# Sparse Time-Frequency Representation for Signals with Fast Varying Instantaneous Frequency

Irena Orović [1], Andjela Draganić [1*], Srdjan Stanković [1]

[1] Faculty of Electrical Engineering, University of Montenegro, Dzordza Vasingtona bb, Podgorica, Montenegro

[*] andjelad@ac.me

**Abstract:** Time-frequency distributions have been used to provide high resolution representation in a large number of signal processing applications. However, high resolution and accurate instantaneous frequency (IF) estimation usually depend on the employed distribution and complexity of signal phase function. To ensure an efficient IF tracking for various types of signals, the class of complex time distributions has been developed. These distributions facilitate analysis in the cases when standard distributions cannot provide satisfactory results (e.g., for highly non-stationary signal phase). In that sense, an ambiguity based form of the forth order complex-time distribution is considered, in a new compressive sensing (CS) context. CS is an intensively growing approach in signal processing that allows efficient analysis and reconstruction of randomly under-sampled signals. In this paper, the randomly chosen ambiguity domain coefficients serve as CS measurements. By exploiting sparsity in the time-frequency plane, it is possible to obtain highly concentrated IF using just small number of random coefficients from ambiguity domain. Moreover, in noisy signal case, this CS approach can be efficiently combined with the L-statistics producing robust time-frequency representations. Noisy coefficients are firstly removed using the L-statistics and then reconstructed by using CS algorithm. The theoretical considerations are illustrated using experimental results.

## 1. Introduction

Highly-localized signal IF in the time-frequency (TF) plane has attracted researchers attention and has been widely studied in the literature [1]-[6]. In order to deal with different types of signals, a number of TF distributions is developed. The quadratic distributions (Wigner-Ville distribution – WD and distributions from the Cohen class) can perfectly localize linear frequency modulated signals. If the considered signal is multicomponent, the WD introduces undesired components called cross-terms. Therefore, in order to reduce or eliminate the cross-terms in the WD, distributions from the Cohen class are employed [5]-[7]. These are based on low-pass filtering of the ambiguity function, which is two-dimensional Fourier transform of the WD. Having auto terms around the origin and cross-terms dislocated from the origin in the ambiguity plane, the resulted TF distribution will have reduced number of cross-terms, or will be cross-terms free. However, in general there is a trade-off between localization in the TF plane and cross-terms reduction, which is known as TF uncertainty principle [4].

Signals, such as radar signals, or vibrating tones of musical instruments, [8], [9] are characterized by the fast varying IF. Analysis of such signals using standard distributions (such as WD and Cohen class distributions) is difficult. In some cases it is even impossible to detect accurately IF changes. Therefore, complex-time distributions were introduced [10]-[14] to highly concentrate phase derivatives in the TF plane and, consequently, provide the exact IF tracking. The complex-time distributions of different orders [3], [11], are applicable to various types of signals, depending on the phase function. The focus of this paper will be on the fourth order complex-time distribution, based on ambiguity domain realization. Most of the real signals exhibit sparsity property in certain domain (time, frequency, time-frequency, etc.) [1], [2], [23]. Sparsity means that in certain domain, the signal could be represented with small number of non-zero coefficients. Sparse signals are the subject of CS application, which is a new and intensively studied topic in signal processing [15]-[32]. CS uses signals sampled at the rates below Nyquist, and provides successful signal reconstruction using small set of randomly selected samples. Random sampling is necessary condition that should be satisfied, in order to reconstruct signal from incomplete set of samples. The majority of real signals is sparse in one domain, while they are dense in another. Here, we deal with signals that are sparse in the joint TF domain, i.e., signals whose IF occupies only small part of the TF plane [1], [2], [23], [24]. By exploiting sparsity in the TF domain, the CS approach is applied to the signal in order to provide better IF localization with reduced number of samples.

It was shown that the ambiguity function can be combined with CS to provide sparse TF representation especially for linear IF [23], [24]. The observations are used from the ambiguity domain, usually from a priori defined area. The shape and size of the area is defined by the mask that should be large enough to collect auto-terms and small enough to avoid the cross-terms. Here we deal with much complex signal structure, and thus we explore the use of ambiguity based complex-time distribution in order to provide sparse and cross-terms free representation and to facilitate samples acquisition. In the cases of noisy signals, the robust form is provided using the L-statistics.

The paper is organized as follows. The theoretical background on commonly used TF distributions is given in the Section II. Distributions with the complex-time argument are described in this section, as well. Section III presents the CS approach in the TF domain. Robust approach to the IF estimation using complex-time distributions and CS approach is described in Section IV. Section V contains several examples which justify presented theory. Conclusion is given in Section VI.

## 2. A review of time-frequency distributions

Real signals differ in their nature: stationarity, sparsity, number of components present in the signal, etc. Therefore, different tools and method have been used for signal processing and analysis. The target signals in this paper are non-stationary signals with fast oscillations of spectral content, which entail TF based approaches.

Generally speaking, an ideal TF distribution should concentrate energy along IF in the TF plane. If we consider the signal in the form $Ae^{j\phi(t)}$, where $A$ is the signal amplitude and $\phi(t)$ denotes the phase, the ideal TF representation should be in the form [3]:

$$IF_{TF}(t,\omega) = 2\pi A^2 \delta(\omega - \phi'(t)). \qquad (1)$$

However, in the real cases there is an energy spreading around IF that appears as a consequence of higher order phase derivatives. The spreading term depends on the signal nature as well as the chosen TF distribution. Therefore, a more realistic form of TF distribution includes the spreading factor as follows:

$$TF(t,\omega) = 2\pi A^2 \delta(\omega - \phi'(t)) *_\omega \mathcal{F}\{w(t)\} *_\omega \mathcal{F}\{e^{jS(t,\tau)}\}, \qquad (2)$$

where $S(t,\tau)$ denotes spreading factor, $w(t)$ is the window function and $\mathcal{F}$ denotes Fourier transform. One of the oldest and simplest TF distributions is spectrogram (quadratic version of the short-time Fourier transform):

$$SPEC(t,\omega) = |STFT(t,\omega)|^2 = \left|\int_{-\infty}^{\infty} x(t+\tau)w(\tau)e^{-j\omega\tau}d\tau\right|^2, \qquad (3)$$

where $x(t)$ is the signal and $w(t)$ is the window function. Spread factor for the spectrogram (obtained using Taylor series expansion) contains all orders of phase derivatives higher than the first one: $S(t,\tau) = \sum_{i=2}^{\infty} \phi^{(i)}(t)\tau^i / i!$. Also, the spectrogram resolution in the TF plane is window dependent. Hence, usually the quadratic distributions are used to overcome this drawback. Commonly used quadratic TF distribution is the WD defined as:

$$WD(t,\omega) = \int_{-\infty}^{\infty} x(t+\frac{\tau}{2})x^*(t-\frac{\tau}{2})e^{-j\omega\tau}d\tau, \qquad (4)$$

It ideally concentrates linear frequency modulated signals, with the zero spread factor. However, in the cases of nonlinear IF changes, the WD produces unwanted inner-interference terms. Spread factor will contain all odd phase derivatives: $S(t,\tau) = \sum_{i=1}^{\infty} \phi^{(2i+1)}(t)(\tau/2)^{2i+1} / (2i+1)!$. Also, when deals with multicomponent signals, the WD produces cross-terms. Cohen class distributions, the S-method, and the distributions with "complex-time argument", are introduced with the aim to overcome drawbacks of the WD. We will focus on the ambiguity domain based distributions, i.e. the Cohen class and class of "complex-time" distributions. The ambiguity function (AF) is defined as a two-dimensional Fourier transform of the WD:

$$A(\theta,\tau) = \mathcal{F}_{t,\omega}\{WD(t,\omega)\} = \int_{-\infty}^{\infty} x(t+\frac{\tau}{2})x^*(t-\frac{\tau}{2})e^{-j\theta t}dt. \tag{5}$$

Having in mind that signal terms are located around the origin in the ambiguity plane and the cross-terms are dislocated, they can be easily reduced by applying the low-pass kernel function. It leads to the distributions from the Cohen class defined as follows:

$$CD(t,\omega) = \frac{1}{2\pi}\int_{-\infty}^{\infty}\int_{-\infty}^{\infty} K(\theta,\tau)A(\theta,\tau)e^{-j\theta t - j\omega\tau}d\tau d\theta, \tag{6}$$

where $K(\theta,\tau)$ denotes 2D kernel function. Choosing an appropriate kernel function and adjusting its parameters, it is possible to reduce or completely remove the cross-terms.

However, in the cases of the signal with fast varying IF, the Cohen class distributions cannot provide satisfactory results in tracking the IF changes. Therefore, higher order distributions based on the complex-time argument should be used to provide high concentration in the TF plane, and consequently, better IF estimation for signals with fast varying IF.

A General form of the complex-time distribution of *N*-th order is defined as:

$$CTD_N(t,\omega) = \int_{-\infty}^{\infty} R_N(t,\tau)e^{-j\omega\tau}d\tau, \tag{7}$$

where $R_N(t,\tau)$ is the complex-time moment, defined as:

$$R_N(t,\tau) = \prod_{k=0}^{N-1} x^{e^{-\frac{j2\pi k}{N}}}\left(t+\frac{\tau}{N}e^{\frac{j2\pi k}{N}}\right), \tag{8}$$

Or,

$$R_N(t,\tau) = \prod_{\substack{i=-N/2,\\ i\neq 0}}^{N/2} x(t+\frac{\tau}{Nsign(i)(a_i+jb_i)})^{sign(i)(a_i+jb_i)}, \tag{9}$$

where *N* is an even number which represents the distribution order, while $a_i$ and $b_i$ denote points on the unit circle. The fourth order complex-time distribution is obtained by using *N*=4:

$$CTD_4(t,\omega) = \int_{-\infty}^{\infty} R_4(t,\tau)e^{-j\omega\tau}d\tau = \int_{-\infty}^{\infty}\prod_{\substack{i=-2,\\ i\neq 0}}^{2} x(t+\frac{\tau}{4sign(i)(a_i+jb_i)})^{sign(i)(a_i+jb_i)}e^{-j\omega\tau}d\tau, \tag{10}$$

where the spread factor can be defined in the form:

$$S(t,\tau) = \phi^{(1)}(t)\tau + \phi^{(3)}(t)\frac{\tau^3}{16\cdot 3!(a+jb)^2} + \phi^{(5)}(t)\frac{\tau^5}{256\cdot 5!(a+jb)^4} + \phi^{(7)}(t)\frac{\tau^7}{4096\cdot 7!(a+jb)^6} + \ldots \tag{11}$$

Starting from the moment function with the complex-time argument, it is possible to define the real-time and complex-time ambiguity functions. The following relation can be referred as complex-time ambiguity function [3]:

$$A_{ct}(\theta,\tau) = \int_{-\infty}^{\infty} R_{c4}(t,\tau)e^{-j\theta t}dt, \quad (12)$$

where $R_{c4}(t,\tau)$ denotes imaginary part of the moment function $R_4(t,\tau)$. The real-time ambiguity function could be defined, in vector form, as [3]:

$$A_{rt}(\theta,\tau) = \int_{-\infty}^{\infty} x(t+\frac{\tau}{N})x^*(t-\frac{\tau}{N})e^{-j\theta t}dt. \quad (13)$$

Both of the ambiguity functions (12) and (13) can be kernel filtered:

$$\begin{aligned}A_{rt}^K(\theta,\tau) &= K(\theta,\tau)A_{rt}(\theta,\tau),\\ A_{ct}^K(\theta,\tau) &= K(\theta,\tau)A_{ct}(\theta,\tau),\end{aligned} \quad (14)$$

The resulting ambiguity function could be obtained by convolving with the window function $W(\varepsilon)$ as follows [3]:

$$A_{CTD}(\theta,\tau) = \int_{-\infty}^{\infty}\int_{-\infty}^{\infty}\int_{-\infty}^{\infty} W(\varepsilon)e^{-j\varepsilon\tau_1}e^{j\varepsilon(\tau-\tau_1)}A_{rt}^K(\theta_1,\tau_1)A_{ct}^K(\theta-\theta_1,\tau-\tau_1)d\tau_1 d\theta_1 d\varepsilon, \quad (15)$$

Now, complex-time distribution could be obtained as:

$$CTD(t,\omega) = \frac{1}{2\pi}\int_{-\infty}^{\infty}\int_{-\infty}^{\infty} A_{CTD}(\theta,\tau)e^{j\theta t - j\omega\tau}d\tau d\theta. \quad (16)$$

By properly choosing complex-time distribution order, higher order phase derivatives can be decreased and distribution concentration can be improved.

In the sequel, we propose an approach to combine (and take advantage of) the TF distributions with complex-time argument and Compressive Sensing approach. This combination can be exploited to provide sparse TF distribution and efficient IF estimation.

## 3. Time-Frequency representations and Compressive Sensing

### a. *Compressive Sensing concept*

Compressive Sensing (CS) [15]-[20] is widely studied approach in the recent years. It provides successful signal reconstruction, using incomplete set of signal samples, i.e. deals with signals sampled at the rate lower than Nyquist. Signal can be intentionally under-sampled. In the noisy signal cases, corrupted samples can be considered as missing ones, if we are able to locate these samples in the signal. In order to apply CS procedure, certain conditions such as sparsity and incoherence, should be satisfied. Sparsity refers to the signal property which states that it has to exist transform domain in which signal can be represented with a small number of non-zero samples. Incoherence provides successful reconstruction with available small number of signal samples. Most of the real signals satisfy these two conditions, which means that the area of possible CS applications is wide.

The signals can be reconstructed from a small number of available samples using complex optimization algorithms, if the sparsity property is satisfied. Assume that we have time domain signal $x$, of length $N$. Let $\mathcal{F}^{-1}$ be sparsifying basis for the signal $x$. Therefore, signal can be represented in form of the basic matrix as follows:

$$x = \mathcal{F}^{-1}_{N \times N} S. \tag{17}$$

If $\mathcal{F}^{-1}_{N \times N}$ denotes $N \times N$ inverse Fourier transform matrix, then $S$ denotes vector of Fourier transform coefficients. In the matrix form, previous relation becomes:

$$\begin{bmatrix} x(0) \\ x(1) \\ \ldots \\ x(N-1) \end{bmatrix} = \begin{bmatrix} 1 & 1 & \ldots & 1 \\ 1 & e^{j\frac{2\pi}{N}} & \ldots & e^{j\frac{2(N-1)\pi}{N}} \\ \ldots & \ldots & \ldots & \ldots \\ 1 & e^{j\frac{2(N-1)\pi}{N}} & \ldots & e^{j\frac{2(N-1)(N-1)\pi}{N}} \end{bmatrix} \begin{bmatrix} S(0) \\ S(1) \\ \ldots \\ S(N-1) \end{bmatrix}, \tag{18}$$

where each element of the matrix $\mathcal{F}^{-1}_{N \times N}$ is exponential term $e^{j(2\pi kn/N)}$, $k = 0, \ldots, N-1; n = 0, \ldots, N-1$. Assume now that only $M$ randomly distributed samples of the signal $x$ are known, while the rest ($N$-$M$) samples are considered as missing samples. This means that Fourier transform matrix is not full $N \times N$ matrix. It is randomly subsampled matrix, and contains $M$ out of $N$ randomly selected rows or the columns of the original matrix. Random selection can be modeled as matrix multiplication of the original matrix $\mathcal{F}^{-1}_{N \times N}$ with the incoherent measurement matrix $\Phi_{M \times N}$. Matrix formed in such way is called random partial Fourier matrix, and can be described with following relation:

$$\mathcal{F}^{-1}_{P} = \mathcal{F}^{-1}_{N \times N} \Phi_{M \times N} = \begin{bmatrix} 1 & 1 & \ldots & 1 \\ 1 & e^{j\frac{2\pi}{N}} & \ldots & e^{j\frac{2\pi}{N}(N-1)} \\ \ldots & \ldots & \ldots & \ldots \\ 1 & e^{j\frac{2\pi}{N}M} & \ldots & e^{j\frac{2\pi}{N}M(N-1)} \end{bmatrix}. \tag{19}$$

Randomly chosen $M$ rows (columns) of the matrix $\mathcal{F}_{N\times N}^{-1}$ results in the measurement vector $v$. Measurement vector is defined as:

$$v = \begin{bmatrix} 1 & 1 & \cdots & 1 \\ 1 & e^{j\frac{2\pi}{N}} & \cdots & e^{j\frac{2\pi}{N}(N-1)} \\ \cdots & \cdots & \cdots & \cdots \\ 1 & e^{j\frac{2\pi}{N}M} & \cdots & e^{j\frac{2\pi}{N}M(N-1)} \end{bmatrix} s, \tag{20}$$

where $\mathcal{F}_P^{-1}$ is the CS matrix. Relation (20) represents an undetermined system of equations. In order to find the sparsest solution for this system of equations, among large number of possible solutions, optimization algorithms are used. The sparsest solution can be obtained using variety of algorithms, such as greedy algorithms (MP, OMP, StOMP, CoSaMP, etc.) [25]-[28], convex relaxation algorithms and the least absolute shrinkage and selection operator (LASSO) [29], non-iterative and iterative solutions, etc. Commonly used optimization algorithms are based on $\ell_1$-norm minimization [17]:

$$\min_s \|s\|_1 \text{ subject to } \quad v = \mathcal{F}_P^{-1} s, \tag{21}$$

### b. Extended problem formulation – 2D partial Fourier transform matrix

The CS approach could be applied to the TF domain in order to provide sparse representation and better localization of the signal in the TF plane [23], [24]. When considering TF domain as domain of sparsity, the ambiguity domain is used as a domain of observation. Therefore, measurements are randomly selected from the ambiguity domain. Having in mind the relation between ambiguity domain and the WD:

$$A(\theta,\tau) = \mathcal{F}^{2D} \cdot WD(t,\omega), \tag{22}$$

The following relation can be derived:

$$A(\theta,\tau) = \mathcal{F}^{2D} \cdot CTD(t,\omega), \tag{23}$$

where $CTD$ denotes complex time distribution, and $\mathcal{F}_{2D}$ is 2D $N^2 \times N^2$ Fourier transform matrix. Therefore, here, the dense counterpart of the ambiguity domain is 4th order complex time distribution $CTD$. 2D FT is produced as Kronecker product of the identity matrix $I$ and 1D FT matrix:

$$\mathcal{F}_{N^2 \times N^2}^{2D} = I_{N\times N} \otimes \mathcal{F}_{N\times N}^{1D}, \tag{24}$$

where $\otimes$ denotes Kronecker product. In the matrix form:

$$\begin{bmatrix} \mathcal{F}^{1D}_{N\times N} & 0 & ... & 0 \\ 0 & \mathcal{F}^{1D}_{N\times N} & ... & 0 \\ ... & ... & ... & 0 \\ 0 & 0 & 0 & \mathcal{F}^{1D}_{N\times N} \end{bmatrix} = \begin{bmatrix} 1_{N\times N} & 0 & ... & 0 \\ 0 & 1_{N\times N} & ... & 0 \\ ... & ... & ... & 0 \\ 0 & 0 & 0 & 1_{N\times N} \end{bmatrix} \otimes \begin{bmatrix} 1 & 1 & ... & 1 \\ 1 & e^{-j\frac{2\pi}{N}} & ... & e^{-j\frac{2(N-1)\pi}{N}} \\ ... & ... & ... & ... \\ 1 & e^{-j\frac{2(N-1)\pi}{N}} & ... & e^{-j\frac{2(N-1)^2\pi}{N}} \end{bmatrix}. \quad (25)$$

Let us describe ambiguity domain based CS problem. For the $N\times N$ TF representation, at most $K\times N$ non-zero components should exist, where $K$ is number of signal components [23]. If we consider only the coefficients around the origin in the ambiguity plane, and use these coefficients as measurements for the CS procedure, obtained ambiguity function is measurement ambiguity function. It can be described as follows:

$$A^M_{S^2\times 1} = \begin{bmatrix} 0_{L\times L} & 0_{L\times S} & 0_{L\times L} \\ 0_{S\times L} & 1_{S\times S} & 0_{S\times L} \\ 0_{L\times L} & 0_{L\times S} & 0_{L\times L} \end{bmatrix} \begin{bmatrix} \mathcal{F}^{1D}_{N\times N} & 0 & ... & 0 \\ 0 & \mathcal{F}^{1D}_{N\times N} & ... & 0 \\ ... & ... & ... & 0 \\ 0 & 0 & 0 & \mathcal{F}^{1D}_{N\times N} \end{bmatrix} A_{N^2\times 1} = \mathcal{F}^{\Lambda}_{S^2\times N^2} A_{N^2\times 1}, \quad (26)$$

where matrix $\mathcal{F}^{\Lambda}_{S^2\times N^2}$ denotes 2D partial Fourier transform matrix. Sparse TF distribution can be obtained by minimizing the following function:

$$\min_{\sigma}\left(G(\sigma)+F(\sigma)\right), \quad (27)$$

where

$$\begin{aligned} G(\sigma) &= \lambda\|\sigma\|_1, \\ F(\sigma) &= (A^M - \mathcal{F}^{\Lambda}\cdot\sigma)/2 \end{aligned}, \quad (28)$$

and σ is estimated sparse TF distribution.

## 4. Robust approach to the IF estimation

In the cases when the ambiguity function is corrupted with impulsive noise, the randomly selected coefficients used as measurements in CS procedure, will be noisy as well. The initial transform domain vector, used in CS optimization problem, should be noise free. Therefore, the robust statistics can be applied to the coefficients of the ambiguity function in order to remove noisy peaks. This is achieved by using L-statistics that perform well in the presence of impulsive and mixed noise [23], [24]. According to the L-estimation approach, the noisy measurement vector is sorted, and certain percent of the coefficients is discarded – both the smallest and largest value coefficients. Therefore, the L-estimation based minimization problem can be defined as follows:

$$\arg\min_{s} \|s\|_1 \quad \text{subject to} \quad v = \mathcal{F}_P^{-1} s. \tag{29}$$

Robust initial transform is obtained as [23]:

$$s_0 = \sum_{i=P}^{Q} AF_{SORT},$$
$$AF_{SORT}(\tau,\theta) = sort\{AF(\tau,\theta)e^{-j2\pi\tau k/M}e^{-j2\pi\theta l/M}\}. \tag{30}$$

where $P$ and $Q$ denote number of discarded smallest and largest value coefficients, respectively.

## 5. Experimental results

### Example 1: Simulated radar signal with fast varying IF changes

Consider the multicomponent signal with the nonlinear phase function in the form:

$$x(t) = e^{j(2\cos(\pi t)+\cos(4\pi t)+4.5\pi t)/2} +$$
$$+ e^{j(\cos(\pi t)+\cos(3\pi t)+\cos(4\pi t)-8\pi t)/2}, \tag{31}$$

The first phase derivative, which corresponds to the signal IF, is shown in the Figure 1a.

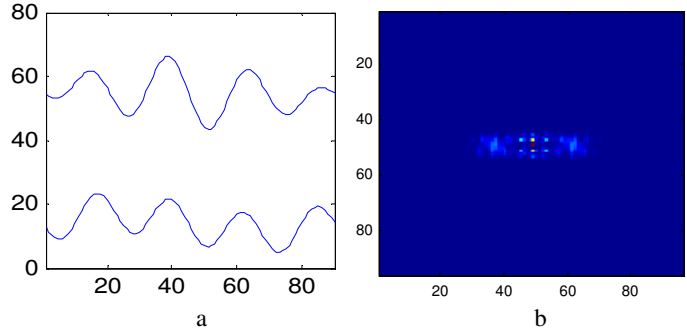

*Fig. 1.* a. The first phase derivative of the signal (31) and its b. ambiguity function

In order to find suitable TF representation for the observed signal, several TF representations are considered. The TF representations are of 90×90 size and they are displayed in Figure 2. The IF estimated from the observed distributions are shown with red line, while the correct IF is shown with blue line. The time-frequency mask is formed by using central region of 25×25 size in the ambiguity domain. This region contains 7.7% of the total number of samples. The 50% of the samples from the mask (3.8% of the total number of samples) are chosen randomly and serve as measurements in the CS procedure.

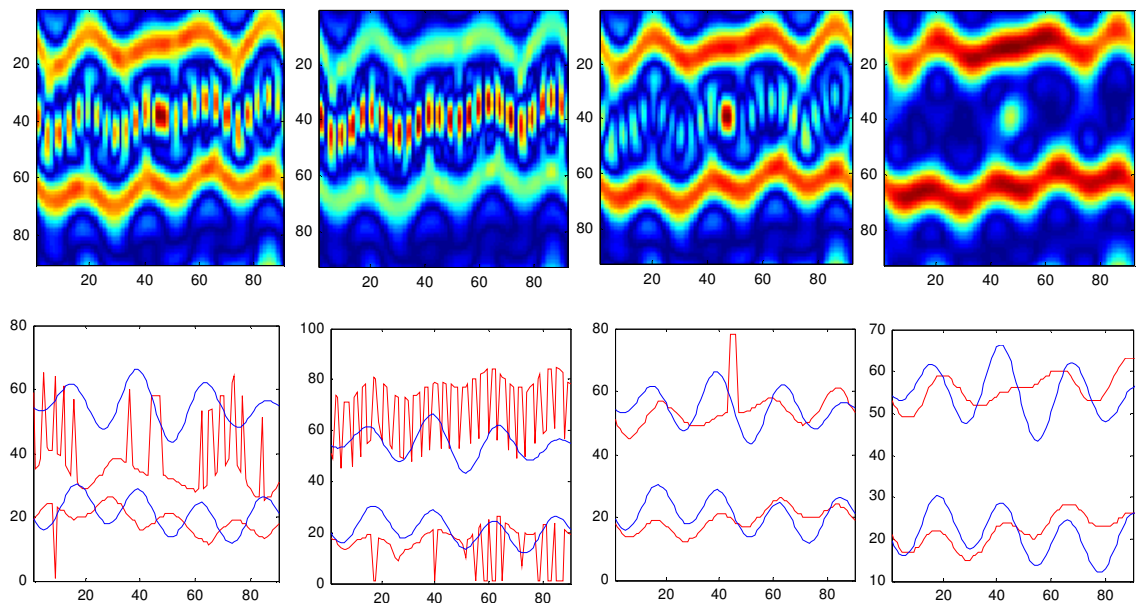

*Fig. 2. 1st row, from left to right: the WD of the signal (31), distribution from the Cohen, obtained by using Gaussian kernel with different parameters; 2nd row - corresponding IF estimated from the WD and distribution from the Cohen class obtained by using Gaussian kernel with different parameters. Blue line is for correct IF and red line corresponds to the estimated IF*

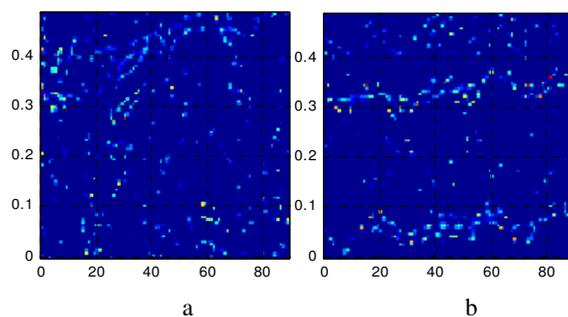

*Fig. 3. Sparse TF distribution obtained from the: a. the WD, b. the Cohen class distribution based on Gaussian kernel*

Firstly, the WD distribution is calculated. Apart from the inability to capture fast changes of the IF, the WD introduces cross-terms in the TF representation. Therefore, different Cohen class distributions are tested on the same signal, in order to provide cross-terms free representation. Distributions from the Cohen class enable controlling the amount of cross-terms, by changing kernel functions and their parameters, but fail to provide accurate IF tracking of the non-stationary signals, as it is shown in the figures.

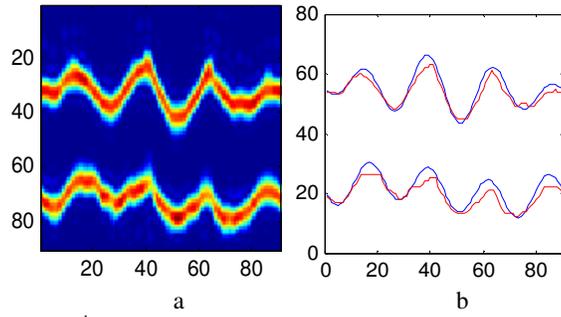

***Fig. 4.*** *a. $4^{th}$ order complex-time distribution of the signal (31);*
*b. Red line: IF of the signal (31) estimated form the distribution in Fig 4a, blue line-correct IF of the signal*

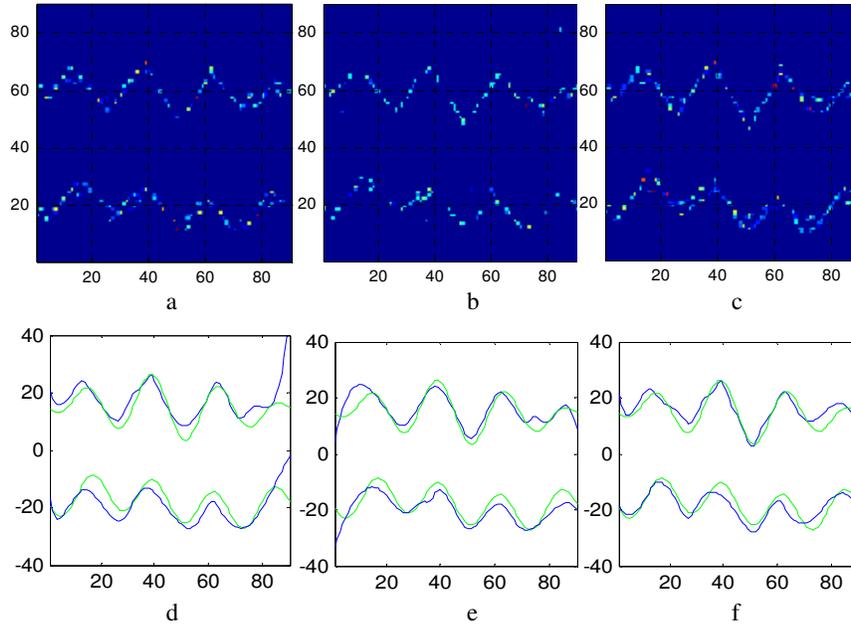

***Fig. 5.*** *Sparse time-frequency representations obtained by using:*

*a. 40%, b. 50%, c. 60% of randomly chosen measurements from 25×25 mask;*

*d.; e.; and f. : IF frequency obtained from the corresponding distributions a), b) and c); blue line is for estimated IF and green line is for correct IF*

Also, there is a trade-off between cross-terms reduction and concentration in the TF plane. The WD and distribution from the Cohen class are shown in the Figure 2. As a kernel function, in all considered cases the Gaussian low pass filtering function is used. Parameters of the Gaussian kernel are changed, and consequently, several TF distributions are obtained. Both, the WD and the Cohen class distribution, fail to track IF changes accurately. Also, cross-terms are present in both cases. Figure 3 shows resulting sparse time-frequency distributions, obtained by using the WD and the Cohen class distribution. Note that the resulting distributions are not sparse and does not track accurately IF changes. The complex-time distribution of the fourth order, calculated for the signal (31), is shown in Figure 4. The ambiguity domain filtering with Gaussian kernel is used. The mask size is 25×25. As it can be seen from the Figure 5, the resulting sparse TF distribution is provided using 40% of the samples from the mask, which is only 3% of the total number of samples. Also, the resulting IF corresponds to the real IF of the signal.

**Table 1** MSE of the IF estimation

| Distribution | MSE | |
|---|---|---|
| | Comp 1 | Comp 2 |
| Wigner distribution | 3.3192×10³ | 79.7761 |
| Distribution from the Cohen class based on Gaussian kernel $e^{-(\tau^2+\theta^2)/\delta^2}$ with $\delta$=120 | 1.1563×10³ | 3.1624×10³ |
| Distribution from the Cohen class based on Gaussian kernel $e^{-(\tau^2+\theta^2)/\delta^2}$ with $\delta$=80 | 1.3880×10³ | 1.5905e×10³ |
| Distribution from the Cohen class based on Gaussian kernel $e^{-(\tau^2+\theta^2)/\delta^2}$ with $\delta$=20 | 1.6179×10³ | 1.7346×10³ |
| *Forth order complex time distribution:* | | |
| 7×7 mask and all samples within the mask used in CS procedure | failed | Failed |
| 10×10 mask and all samples within the mask used in CS procedure | failed | Failed |
| 15×15 mask and 70% samples used in CS procedure | 81.3099 | 57.1962 |
| 20×20 mask and 60% samples used in CS procedure | 8.9314 | 20.5861 |
| *25×25 mask* | | |
| 40% samples from the mask | 17.7834 | 32.7570 |
| 50% samples from the mask | 10.7177 | 9.2834 |
| 60% samples from the mask | 7.3854 | 8.1211 |

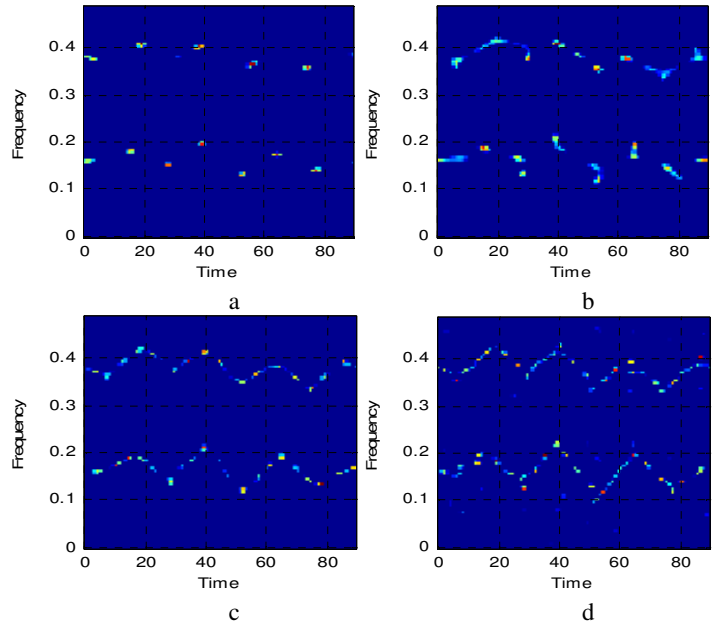

***Fig. 6.*** *Sparse time-frequency distributions obtained using different mask sizes and different percent of used samples:*

*a. 7×7 mask with 100% of samples;*

*b. 10×10 mask with 100% of samples;*

*c. 15×15 mask with 70% of samples;*

*d. 20×20 mask with 60% of samples*

Let us consider mask size influence on the number of required samples for providing sparse TF representation. As it can be seen from the Figure 6, 7×7 and 10×10 masks are too small to provide accurate IF representation. Larger masks (15×15 and 20×20) can provide an accurate IF estimation, but require larger number of samples to be used as CS measurements (i.e. 70% and 60% of the samples from the mask has to be used, resprectively) in order to obtain an accurate IF estimation.

*Example 2: Noisy signal measurements*

Let us now consider simulated radar signal. Monocomponent signal is periodically frequency modulated and has fast IF changes. Nonuniform rotation of the reflecting point in radar systems could be described with this form of signal [8]. Signal is of 96 seconds duration, sinusoidally modulated and it is sampled at a frequency of 48 Hz: $x(t)=e^{j(4cos(\pi t)+(2/3)cos(3\pi t)+(2/3)cos(5\pi t))}$. Assume that the ambiguity domain measurements are corrupted with impulse type of noise. Original and noisy ambiguity functions are shown in Figure 7a and 7b, respectively. The sparse TF distribution obtained by using 50% of the noisy ambiguity samples from the 25×25 mask, is shown in Figure 7c. Ambiguity function has small number of peaks (Figure 7b), but still enough to disable accurate IF estimation of the observed signal. The obtained CS based TF distribution (Figure 7c) is not sparse and does not provide accurate signal IF. Therefore, the ambiguity function is denoised prior to the measurements selection. The 0.5% of the highest ambiguity function coefficients is removed and robust form of the ambiguity function is obtained (Figure 8a). As it can be seen, noisy peaks are successfully removed and sparse TF with accurate IF tracking is obtained (Figure 8b and 8c).

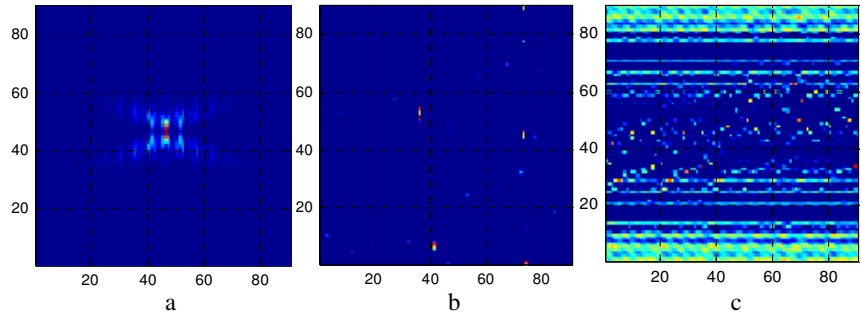

*Fig. 7. a. Original and b. noisy ambiguity function,*
*c. TF distribution obtained using noisy ambiguity measurements*

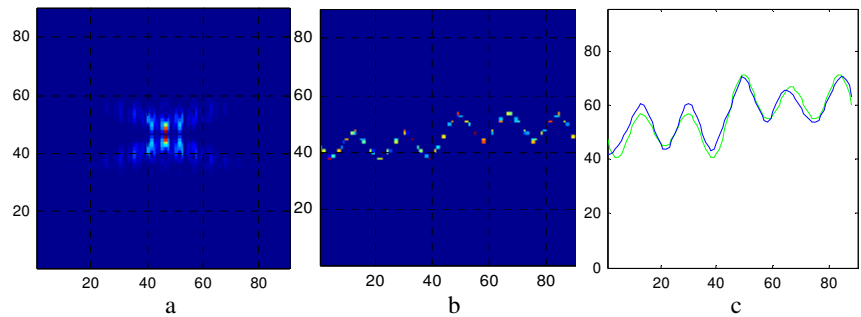

***Fig. 8.*** *a. Robust ambiguity function,*

*b. sparse TF distribution,*

*c. blue-true and green-estimated IF of the signal*

**Example 3: Real radar signal**

Finally, as an example of signal recorded in real noisy environment, let us observe a portion of radar data corresponding to the moving human body. During the walking process, the main body component does not produce strictly linear trajectory (when observed on selected portion), but rather produces the micro movements which can be hardly detected using standard low-resolution TF representations, such as the spectrogram. In that sense the high-resolution solutions provided by the complex-time TF distributions allows more descriptive results, allowing detection of the smallest variations, as shown in Fig. 9.

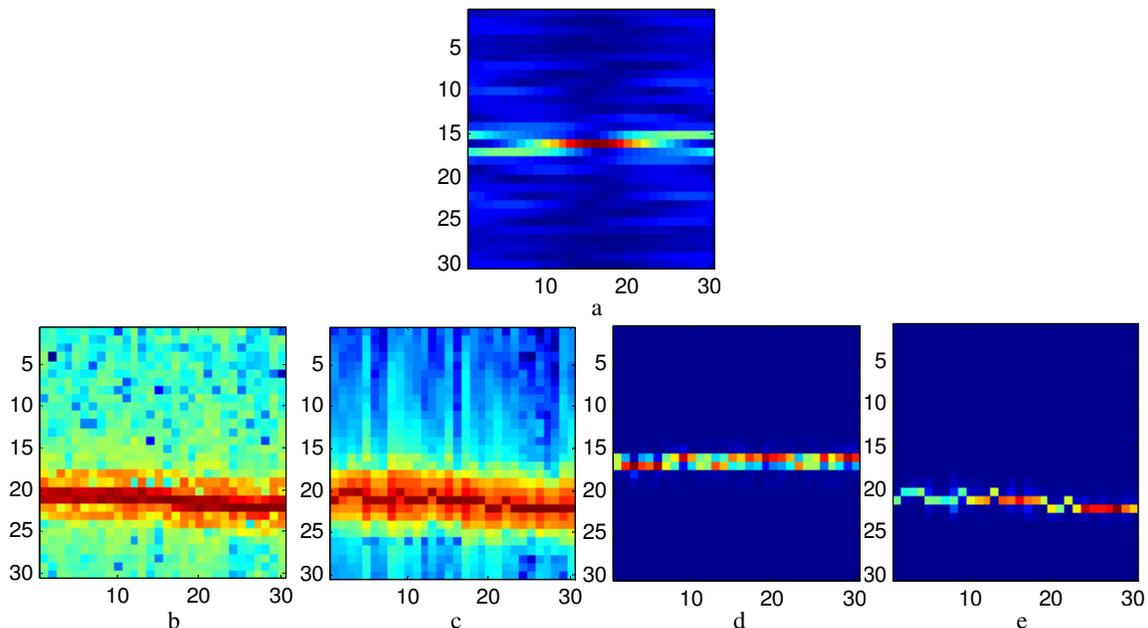

***Fig. 9.*** *a. Ambiguity function of the real radar signal;*

*b. and c. Spectrogram and complex-time distribution of the signal, log scale;*

*d. and e. Spectrogram and complex-time distribution*

These additional movements of the target are called micro-Doppler modulations and contain information about the target which can be used for the target recognition [32]. Obviously, the detected details could be useful for object classification purposes in applications involving object with linear and nonlinear velocity changes.

Now, assume that we are faced with a reduced set of available observations (11% available samples of the total number of samples in TF plane), when we need to employ the proposed CS based approach. The resulted sparse TF distribution is shown in Fig. 10, while the estimated IFs from the original and sparse complex time distributions are shown in Fig. 11a and Fig. 11b. As it can be seen from the Fig. 11, the IF estimated form the sparse TF corresponds to the IF estimated form the original complex time distribution.

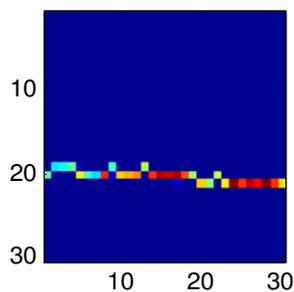

*Fig. 10. Sparse complex-time distribution*

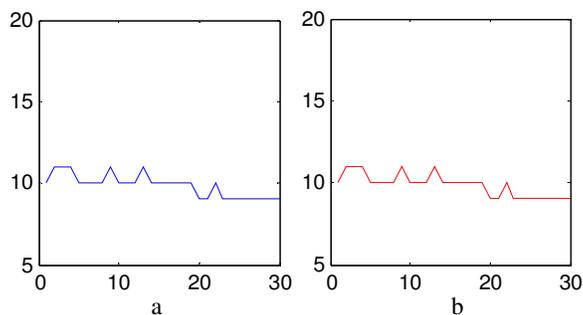

*Fig 11. Estimated IFs: a. from the original complex-time;
b. from the sparse complex-time distribution*

## 6. Conclusion

This paper deals with the CS based IF estimation from the sparse TF representation. Highly non-stationary signals are observed. Having in mind that standard distributions fail to estimate IF of the highly non-stationary signals accurately, in this paper we have used the complex time distributions. These distributions are able to provide TF suitable for IF estimation. The observations used in CS procedure are taken randomly from the ambiguity domain. Depending on the signal structure, certain percent of the measurements are taken from the region around the origin or from the whole ambiguity domain. Signals whose ambiguity function is concentrated around the origin are reconstructed and IF is estimated by using around 4% of samples from the ambiguity domain. The IF estimation accuracy is measured with MSE.